\documentclass[twocolumn,letterpaper,aps,prl,superscriptaddress,showpacs,floatfix]{revtex4}

\usepackage{graphicx}	
\usepackage{xspace}     

\begin{document}



\title{Medium modification of jet fragmentation in Au$+$Au
collisions at $\sqrt{s_{_{NN}}}=200$ GeV measured in direct
photon-hadron correlations}

\newcommand{\abilene}{Abilene Christian University, Abilene, Texas 79699, USA}
\newcommand{\acadsin}{Institute of Physics, Academia Sinica, Taipei 11529, Taiwan}
\newcommand{\banaras}{Department of Physics, Banaras Hindu University, Varanasi 221005, India}
\newcommand{\barc}{Bhabha Atomic Research Centre, Bombay 400 085, India}
\newcommand{\baruch}{Baruch College, City University of New York, New York, New York, 10010 USA}
\newcommand{\bnlcoll}{Collider-Accelerator Department, Brookhaven National Laboratory, Upton, New York 11973-5000, USA}
\newcommand{\bnlphys}{Physics Department, Brookhaven National Laboratory, Upton, New York 11973-5000, USA}
\newcommand{\caucr}{University of California - Riverside, Riverside, California 92521, USA}
\newcommand{\charlesczech}{Charles University, Ovocn\'{y} trh 5, Praha 1, 116 36, Prague, Czech Republic}
\newcommand{\chonbuk}{Chonbuk National University, Jeonju, 561-756, Korea}
\newcommand{\ciae}{Science and Technology on Nuclear Data Laboratory, China Institute of Atomic Energy, Beijing 102413, P.~R.~China}
\newcommand{\cns}{Center for Nuclear Study, Graduate School of Science, University of Tokyo, 7-3-1 Hongo, Bunkyo, Tokyo 113-0033, Japan}
\newcommand{\colorado}{University of Colorado, Boulder, Colorado 80309, USA}
\newcommand{\columbia}{Columbia University, New York, New York 10027 and Nevis Laboratories, Irvington, New York 10533, USA}
\newcommand{\czechtech}{Czech Technical University, Zikova 4, 166 36 Prague 6, Czech Republic}
\newcommand{\dapnia}{Dapnia, CEA Saclay, F-91191, Gif-sur-Yvette, France}
\newcommand{\debrecen}{Debrecen University, H-4010 Debrecen, Egyetem t{\'e}r 1, Hungary}
\newcommand{\elte}{ELTE, E{\"o}tv{\"o}s Lor{\'a}nd University, H - 1117 Budapest, P{\'a}zm{\'a}ny P. s. 1/A, Hungary}
\newcommand{\ewha}{Ewha Womans University, Seoul 120-750, Korea}
\newcommand{\fit}{Florida Institute of Technology, Melbourne, Florida 32901, USA}
\newcommand{\fsu}{Florida State University, Tallahassee, Florida 32306, USA}
\newcommand{\gsu}{Georgia State University, Atlanta, Georgia 30303, USA}
\newcommand{\hanyang}{Hanyang University, Seoul 133-792, Korea}
\newcommand{\hiroshima}{Hiroshima University, Kagamiyama, Higashi-Hiroshima 739-8526, Japan}
\newcommand{\ihepprot}{IHEP Protvino, State Research Center of Russian Federation, Institute for High Energy Physics, Protvino, 142281, Russia}
\newcommand{\illuiuc}{University of Illinois at Urbana-Champaign, Urbana, Illinois 61801, USA}
\newcommand{\inrras}{Institute for Nuclear Research of the Russian Academy of Sciences, prospekt 60-letiya Oktyabrya 7a, Moscow 117312, Russia}
\newcommand{\instpasczech}{Institute of Physics, Academy of Sciences of the Czech Republic, Na Slovance 2, 182 21 Prague 8, Czech Republic}
\newcommand{\isu}{Iowa State University, Ames, Iowa 50011, USA}
\newcommand{\jinrdubna}{Joint Institute for Nuclear Research, 141980 Dubna, Moscow Region, Russia}
\newcommand{\jyvaskyla}{Helsinki Institute of Physics and University of Jyv{\"a}skyl{\"a}, P.O.Box 35, FI-40014 Jyv{\"a}skyl{\"a}, Finland}
\newcommand{\kek}{KEK, High Energy Accelerator Research Organization, Tsukuba, Ibaraki 305-0801, Japan}
\newcommand{\korea}{Korea University, Seoul, 136-701, Korea}
\newcommand{\kurchatov}{Russian Research Center ``Kurchatov Institute", Moscow, 123098 Russia}
\newcommand{\kyoto}{Kyoto University, Kyoto 606-8502, Japan}
\newcommand{\labllr}{Laboratoire Leprince-Ringuet, Ecole Polytechnique, CNRS-IN2P3, Route de Saclay, F-91128, Palaiseau, France}
\newcommand{\lawllnl}{Lawrence Livermore National Laboratory, Livermore, California 94550, USA}
\newcommand{\losalamos}{Los Alamos National Laboratory, Los Alamos, New Mexico 87545, USA}
\newcommand{\lpc}{LPC, Universit{\'e} Blaise Pascal, CNRS-IN2P3, Clermont-Fd, 63177 Aubiere Cedex, France}
\newcommand{\lund}{Department of Physics, Lund University, Box 118, SE-221 00 Lund, Sweden}
\newcommand{\maryland}{University of Maryland, College Park, Maryland 20742, USA}
\newcommand{\mass}{Department of Physics, University of Massachusetts, Amherst, Massachusetts 01003-9337, USA }
\newcommand{\muenster}{Institut fur Kernphysik, University of Muenster, D-48149 Muenster, Germany}
\newcommand{\muhlenberg}{Muhlenberg College, Allentown, Pennsylvania 18104-5586, USA}
\newcommand{\myongji}{Myongji University, Yongin, Kyonggido 449-728, Korea}
\newcommand{\nagasaki}{Nagasaki Institute of Applied Science, Nagasaki-shi, Nagasaki 851-0193, Japan}
\newcommand{\newmex}{University of New Mexico, Albuquerque, New Mexico 87131, USA }
\newcommand{\nmsu}{New Mexico State University, Las Cruces, New Mexico 88003, USA}
\newcommand{\ohio}{Department of Physics and Astronomy, Ohio University, Athens, Ohio 45701, USA}
\newcommand{\ornl}{Oak Ridge National Laboratory, Oak Ridge, Tennessee 37831, USA}
\newcommand{\orsay}{IPN-Orsay, Universite Paris Sud, CNRS-IN2P3, BP1, F-91406, Orsay, France}
\newcommand{\peking}{Peking University, Beijing 100871, P.~R.~China}
\newcommand{\pnpi}{PNPI, Petersburg Nuclear Physics Institute, Gatchina, Leningrad region, 188300, Russia}
\newcommand{\riken}{RIKEN Nishina Center for Accelerator-Based Science, Wako, Saitama 351-0198, Japan}
\newcommand{\rikjrbrc}{RIKEN BNL Research Center, Brookhaven National Laboratory, Upton, New York 11973-5000, USA}
\newcommand{\rikkyo}{Physics Department, Rikkyo University, 3-34-1 Nishi-Ikebukuro, Toshima, Tokyo 171-8501, Japan}
\newcommand{\saispbstu}{Saint Petersburg State Polytechnic University, St. Petersburg, 195251 Russia}
\newcommand{\saopaulo}{Universidade de S{\~a}o Paulo, Instituto de F\'{\i}sica, Caixa Postal 66318, S{\~a}o Paulo CEP05315-970, Brazil}
\newcommand{\seoulnat}{Department of Physics and Astronomy, Seoul National University, Seoul, Korea}
\newcommand{\stonybrkc}{Chemistry Department, Stony Brook University, SUNY, Stony Brook, New York 11794-3400, USA}
\newcommand{\stonycrkp}{Department of Physics and Astronomy, Stony Brook University, SUNY, Stony Brook, New York 11794-3400, USA}
\newcommand{\subatech}{SUBATECH (Ecole des Mines de Nantes, CNRS-IN2P3, Universit{\'e} de Nantes) BP 20722 - 44307, Nantes, France}
\newcommand{\tenn}{University of Tennessee, Knoxville, Tennessee 37996, USA}
\newcommand{\titech}{Department of Physics, Tokyo Institute of Technology, Oh-okayama, Meguro, Tokyo 152-8551, Japan}
\newcommand{\tsukuba}{Institute of Physics, University of Tsukuba, Tsukuba, Ibaraki 305, Japan}
\newcommand{\vandy}{Vanderbilt University, Nashville, Tennessee 37235, USA}
\newcommand{\waseda}{Waseda University, Advanced Research Institute for Science and Engineering, 17 Kikui-cho, Shinjuku-ku, Tokyo 162-0044, Japan}
\newcommand{\weizmann}{Weizmann Institute, Rehovot 76100, Israel}
\newcommand{\wigner}{Institute for Particle and Nuclear Physics, Wigner Research Centre for Physics, Hungarian Academy of Sciences (Wigner RCP, RMKI) H-1525 Budapest 114, POBox 49, Budapest, Hungary}
\newcommand{\yonsei}{Yonsei University, IPAP, Seoul 120-749, Korea}
\affiliation{\abilene}
\affiliation{\acadsin}
\affiliation{\banaras}
\affiliation{\barc}
\affiliation{\baruch}
\affiliation{\bnlcoll}
\affiliation{\bnlphys}
\affiliation{\caucr}
\affiliation{\charlesczech}
\affiliation{\chonbuk}
\affiliation{\ciae}
\affiliation{\cns}
\affiliation{\colorado}
\affiliation{\columbia}
\affiliation{\czechtech}
\affiliation{\dapnia}
\affiliation{\debrecen}
\affiliation{\elte}
\affiliation{\ewha}
\affiliation{\fit}
\affiliation{\fsu}
\affiliation{\gsu}
\affiliation{\hanyang}
\affiliation{\hiroshima}
\affiliation{\ihepprot}
\affiliation{\illuiuc}
\affiliation{\inrras}
\affiliation{\instpasczech}
\affiliation{\isu}
\affiliation{\jinrdubna}
\affiliation{\jyvaskyla}
\affiliation{\kek}
\affiliation{\korea}
\affiliation{\kurchatov}
\affiliation{\kyoto}
\affiliation{\labllr}
\affiliation{\lawllnl}
\affiliation{\losalamos}
\affiliation{\lpc}
\affiliation{\lund}
\affiliation{\maryland}
\affiliation{\mass}
\affiliation{\muenster}
\affiliation{\muhlenberg}
\affiliation{\myongji}
\affiliation{\nagasaki}
\affiliation{\newmex}
\affiliation{\nmsu}
\affiliation{\ohio}
\affiliation{\ornl}
\affiliation{\orsay}
\affiliation{\peking}
\affiliation{\pnpi}
\affiliation{\riken}
\affiliation{\rikjrbrc}
\affiliation{\rikkyo}
\affiliation{\saispbstu}
\affiliation{\saopaulo}
\affiliation{\seoulnat}
\affiliation{\stonybrkc}
\affiliation{\stonycrkp}
\affiliation{\subatech}
\affiliation{\tenn}
\affiliation{\titech}
\affiliation{\tsukuba}
\affiliation{\vandy}
\affiliation{\waseda}
\affiliation{\weizmann}
\affiliation{\wigner}
\affiliation{\yonsei}
\author{A.~Adare} \affiliation{\colorado}
\author{S.~Afanasiev} \affiliation{\jinrdubna}
\author{C.~Aidala} \affiliation{\columbia} \affiliation{\losalamos} \affiliation{\mass}
\author{N.N.~Ajitanand} \affiliation{\stonybrkc}
\author{Y.~Akiba} \affiliation{\riken} \affiliation{\rikjrbrc}
\author{R.~Akimoto} \affiliation{\cns}
\author{H.~Al-Bataineh} \affiliation{\nmsu}
\author{H.~Al-Ta'ani} \affiliation{\nmsu}
\author{J.~Alexander} \affiliation{\stonybrkc}
\author{A.~Angerami} \affiliation{\columbia}
% ...
\collaboration{PHENIX Collaboration} \noaffiliation

\date{\today}


\begin{abstract}

The jet fragmentation function is measured with direct 
photon-hadron correlations in $p$$+$$p$ and Au$+$Au collisions at 
$\sqrt{s_{_{NN}}}=200$ GeV. The $p_{T}$ of the photon is an 
excellent approximation to the initial $p_{T}$ of the jet and the 
ratio $z_{T}=p_{T}^{h}/p_{T}^{\gamma}$ is used as a proxy for the 
jet fragmentation function. A statistical subtraction is used to 
extract the direct photon-hadron yields in Au$+$Au collisions while 
a photon isolation cut is applied in $p$$+$$p$. $I_{\rm AA}$, the 
ratio of jet fragment yield in Au$+$Au to that in $p$$+$$p$, 
indicates modification of the jet fragmentation function. 
Suppression, most likely due to energy loss in the medium, is seen 
at high $z_T$. The fragment yield at low $z_T$ is enhanced at large 
angles. Such a trend is expected from redistribution of the lost 
energy into increased production of low-momentum particles.

\end{abstract}

\pacs{25.75.Dw}  

\maketitle


Experiments at the Relativistic Heavy Ion Collider have observed 
the formation of a Quark-Gluon Plasma, reported as a fundamentally 
new state of matter~\cite{whitepaper, starwp, phoboswp, brahmswp}. 
High momentum quarks and gluons (partons) lose energy as they 
traverse this matter, resulting in the observed suppression of high 
transverse momentum (high $p_T$) hadrons in central heavy-ion 
collisions~\cite{ppg003, star, urse, bdmps}.

Direct photons, however, escape the medium 
unmodified~\cite{raa}, since they do not interact via the strong 
force. This makes them an ideal probe with which to calibrate the 
energy of an initial hard scattering. At leading order, direct 
photons are produced via the quantum-chromodynamics analog of 
Compton scattering, $q+g\rightarrow q+\gamma$. In the limit of 
negligible initial transverse momentum, the final state quark and 
photon are emitted back-to-back in azimuth with the photon 
balancing the transverse momentum of the jet arising from the 
quark. Determining the initial momentum of the parton is key to 
measuring the fragmentation function of the quark jet. This initial 
momentum is provided by the measured energy of the unmodified 
direct photon, via direct photon-hadron correlations~\cite{wang}. A 
study of direct photon-hadron correlations in $p$$+$$p$ collisions 
at $\sqrt{s_{_{NN}}} = 200$ GeV has provided a measurement of the 
fragmentation function in agreement with measurements from 
$e^{+}e^{-}$ collisions~\cite{ppg095}. In Au$+$Au collisions 
contributions from next-to-leading-order processes and medium 
induced photon production are expected to be small ($\approx$10\%) 
at high $p_{T}$~\cite{qin}.

Parton energy loss in the medium can be observed as a modification 
to the jet fragmentation function in heavy ion collisions. The 
fragmentation function is defined as $D(z)=\frac{1}{N_{\rm 
jet}}\frac{dN(z)}{dz}$, where $z=p^{h}/p^{\rm jet}$; $p^{\rm jet}$ 
is the initial jet momentum, and $p^h$ is the momentum of a 
hadronic jet fragment. Experimentally, this is accessible using 
direct photon-hadron correlations, where $p^{\gamma}_{T} \approx 
p_{T}^{\rm jet}$.  This balance is only approximate due to the 
transverse momentum, $k_T$, of the colliding partons inside 
nucleons, which on average introduces a transverse momentum 
imbalance and acoplanarity to the photon and its opposing 
jet~\cite{ppg095, ppg039, ppg029}. 

Several energy loss models~\cite{zoww, qin} predicting direct 
photon-hadron correlations only track the medium induced parton 
splitting of the leading parton. Other models follow the lost 
energy, leading to an increase in low momentum (soft) particle 
production. In particular, Borghini and Weidemann~\cite{mlla} use 
the modified leading log approximation (BW-MLLA) and local parton 
hadron duality to first reproduce the measured fragmentation 
function in $e^+e^-$ data. Modeling the energy loss in the medium 
as an increased parton splitting probability, they calculate the 
suppression of high $p_T$ jet fragments, as well as the 
redistribution of energy to lower $p_T$ fragments and resulting 
enhancement at low $z$. The resulting $R_{\rm AA}$ reproduces the 
PHENIX $\pi^{0}$ measurement for 0--10\% central events. The 
yet-another-jet-energy-loss model (YaJEM)~\cite{renk} traces the 
energy lost via gluon radiation and redistribution to soft particle 
production, predicting a suppression of particles at high $z$ and 
an enhancement at low $z$. This calculation has been done 
specifically for $\gamma_{\rm dir}$-$h$, making it directly 
comparable to this data. The predicted low-$z_{T}$ enhancement has 
not yet been observed within the statistical and systematic 
limitations of previously published data~\cite{ppg090, stargjet}.

In this letter, we report fragmentation functions measured in 
Au$+$Au and $p$$+$$p$ collisions determined from the yield of 
hadrons recoiling opposite to direct photons (i.e. the 
``away-side''). The extraction of a purely direct-photon sample is 
complicated by the presence of photons from meson decays 
(dominantly $\pi^{0} \rightarrow \gamma \gamma$), which must be 
removed from the inclusive photon-hadron correlations. PHENIX has 
previously established the extraction of direct photon-hadron 
correlations via a statistical subtraction procedure in 
Au$+$Au~\cite{ppg090} collisions and via an isolation cut in 
$p$$+$$p$ collisions~\cite{ppg095}.

This analysis includes 3.9 billion minimum bias Au$+$Au events 
collected by PHENIX in 2007 and 2.9 billion in 2010, after quality 
cuts. The $p$$+$$p$ data set comprises 0.5 billion photon-triggered 
events collected in 2005 and 2006, corresponding to total recorded 
integrated luminosities of 3.8 (2005) and 10.7 (2006) 
$\rm{pb}^{-1}$, respectively. Details on the $p$$+$$p$ measurement 
were previously presented in~\cite{ppg095}. The kinematic reach and 
improved statistical precision of both data sets allow us to extend 
previous measurements~\cite{ppg090, stargjet}, reaching a low 
momentum fraction, $z\approx 0.1$, where interplay between the 
medium and the deposited energy may be important~\cite{urse}.


The Au$+$Au minimum bias events are triggered by particles firing 
the beam-beam counters (BBCs), which are arrays of \v{C}erenkov 
counters covering 3.1 $< | \eta | <$ 3.9 and 2$\pi$ in azimuth. 
These BBCs are also used to determine the collision centrality and 
the collision vertex position along the beam direction. The 0--40\% 
most central collisions are presented here. Photons and hadrons are 
measured in two central spectrometers spanning $\pi/2$ in azimuth 
and $\pm 0.35$ units of pseudorapidity each~\cite{mainNIM}. The 
photons are measured in one of two electromagnetic 
calorimeters~\cite{emcNIM} and charged hadrons are measured by 
reconstructing tracks in the Drift Chambers and Pad 
Chambers~\cite{trackingNIM}.


In Au$+$Au, a statistical subtraction determines the direct (i.e. 
nondecay) photon-hadron correlations from the measured inclusive 
photon-hadron correlations. Using the measured associated hadron 
yield per inclusive photon, 
$Y_{\rm inc}=1/N_{inc}dN^{h-\gamma_{inc}}/d\Delta\phi$, 
and per decay photon, $Y_{\rm dec}$, the associated yield per 
direct photon, $Y_{\rm dir}$, is determined by~\cite{ppg090}:

\begin{equation} \label{eqn:subtraction}
  Y_{\rm dir} = \frac{R_{\gamma}Y_{\rm inc}-Y_{\rm dec}}{{R_\gamma}-1}.
\end{equation}

\noindent Here $R_\gamma$ is the ratio of inclusive photons to 
decay photons, reported by PHENIX in~\cite{raa}.

Inclusive photon-hadron correlations are determined from the 
distribution of photon-hadron pairs as a function of their 
azimuthal angular separation, $\Delta\phi$. The distribution of 
real pairs is divided by photon-hadron pairs in mixed events to 
correct for the PHENIX acceptance.

The conditional, or per trigger, associated yield is extracted 
after subtraction of photon-hadron pairs from the bulk underlying 
event~\cite{ppg039}. Such particles are expected to be correlated to 
one another
only through the bulk anisotropy of the event, which is 
conventionally
characterized by the Fourier coefficients $v_n$, and 
are removed from the inclusive and decay photon yields using the 
previously measured $v_{2}$~\cite{ppg132}, and neglecting higher 
order terms, according to:

\begin{widetext}
\begin{equation} \label{eqn:jetfunction}
  \frac{1}{N^{t}} \frac{dN^{\rm pair}}{d\Delta\phi} =
   \frac{1}{N^{t}}\frac{N^{\rm pair}_{real}}{\epsilon^{a}\int{d\Delta\phi}}
  \Bigl\{\frac{dN^{\rm pair}_{real}/d\Delta\phi}{dN^{\rm pair}_{\rm mix}/d\Delta\phi}
  - \ b_{0}   \left[ 1 + 2 \langle v^{t}_2 v^{a}_2 \rangle 
   \cos(2\Delta\phi) \right] \Bigr\},
\end{equation}
\end{widetext}

\noindent where the subscripts $t$ and $a$ refer to trigger and 
associated particle, $\epsilon^{a}$ is the detection efficiency for 
the associated particle, and $b_{0}$ indicates the level of 
background pairs. The $\langle v^{t}_2 v^{a}_2 \rangle$ term 
modulates the background rate as a function of $\Delta\phi$, to 
account for correlations arising from flow of the bulk.

The potential effect of ignoring $v_{3}$ in this extraction was 
studied by extrapolating the PHENIX hadron $v_{3}$ 
measurements~\cite{ppg132}. Including a modulation of the 
background by this $v_{3}$ in addition to the $v_{2}$ results in a 
change in the away-side yield on the order of a few percent, 
depending on $\Delta\phi$ and $p_{T}$. The resulting background 
shape uncertainty is minimal for the highest hadron $p_{T}$ 
selections used here due to the low level of combinatorial 
background. The $v_{3}$ effect is included as an additional 
systematic uncertainty on the background subtraction. An absolute 
background normalization is used to fix the background level, 
$b_{0}$, as described in~\cite{ABS}. A {\sc geant} simulation of 
the detector determined the acceptance and efficiency for the 
measured charged hadrons, $\epsilon^{a}$. The uncertainty on 
$\epsilon^{a}$ leads to an 8.8\% overall and 8.0\% normalization 
uncertainty on the yields for the Au$+$Au and $p$$+$$p$ data, 
respectively.

To measure the decay photon contribution, $\pi^{0}$-$h$ 
correlations are constructed following the same method as above 
using $\pi^{0}$s as the trigger. The $\pi^{0}$s are reconstructed 
from photon pairs whose invariant mass is within the window of 
0.12--0.16 GeV/$c^{2}$. The $\pi^{0}$-$h$ correlations are 
translated into decay photon-hadron correlations according to a 
Monte Carlo study of the probability that a $\pi^{0}$ with a given 
$p_{T}$ produces a decay photon within a certain $p_{T}$ bin. This 
procedure is explained in detail in~\cite{ppg090}.

In $p$$+$$p$ collisions, $\gamma_{dir}$-$h$ yields were measured 
using an isolation cut, and removing decay photons from the 
inclusive sample on an event-by-event basis~\cite{ppg095}. In the 
analysis, photons which combine with another photon in the event to 
produce a mass within the $\pi^{0}$ or $\eta$ mass windows were 
rejected. Next, an isolation cut was applied, requiring that the 
transverse electromagnetic energy and charged track momentum within 
a cone of 0.3 rad around the photon be less than 10\% of the photon 
energy. Finally, a statistical subtraction similar to that in the 
Au$+$Au analysis eliminated contributions from decay photons which 
appeared isolated or whose decay partner was lost due to finite 
detector acceptance or efficiency. In $p$$+$$p$ collisions, the 
underlying event is subtracted assuming the yield of photon-hadron 
coincidences is zero at the minimum point in the correlation as a 
function of $\Delta\phi$ (ZYAM). The lowest three points outside 
the isolation cut region are averaged; as there is no flow in 
$p$$+$$p$ collisions, the background is assumed to be flat in 
$\Delta\phi$. By first eliminating photons from other sources 
event-by-event, the signal to background ratio is improved and 
final uncertainties are reduced. The $p$$+$$p$ results using the 
isolation cut agree with a statistical subtraction analysis in 
$p$$+$$p$, but have smaller uncertainties. The high multiplicity of 
the underlying event makes it difficult to perform an isolation cut 
in Au$+$Au, so a statistical subtraction procedure is used instead.


\begin{figure}[htbp]
\centering
\includegraphics[width=1.0\linewidth]{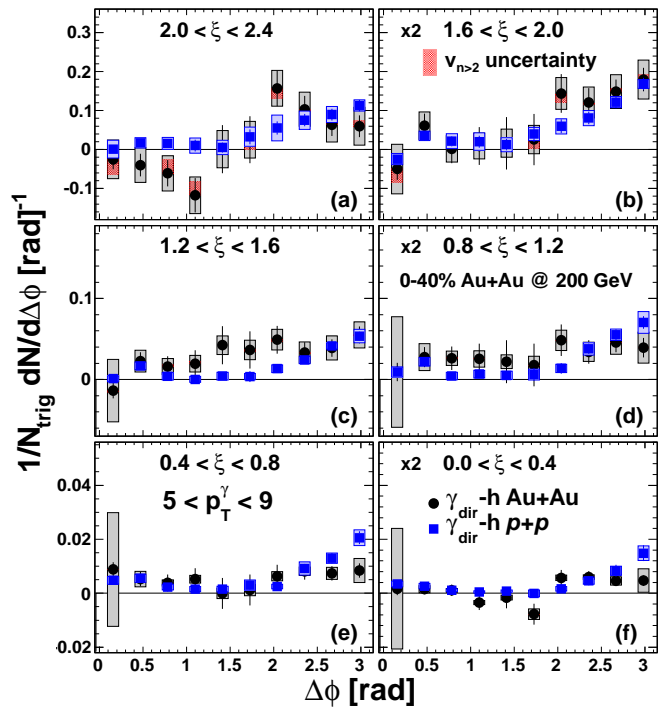}
\caption{(Color online) 
$\Delta\phi$ distribution for various associated $\xi$ bins for 
direct photons (circles) for the 0--40\% most central Au$+$Au collisions and the 
$p$$+$$p$ reference (squares) in all panels. Panels (b), (d), and 
(f) are multiplied by a factor of 2 as indicated. }
\label{fig:dphifig1}
\end{figure}

In order to study the jet fragmentation function, $D(z)$, 
associated hadron yields are determined as a function of 
$z_{T}=p_{T}^{h}/p_{T}^{\gamma}$, the ratio of the associated 
hadron transverse momentum, $p_{T}^{h}$, to the trigger photon 
transverse momentum, $p_T^{\gamma}$. Here $z_T \approx z$, since 
direct photon triggers balance the opposing jet. To focus on the 
low $z_{T}$ region, one can express the fragmentation function as a 
function of the variable, $\xi=ln(1/z_{T})$. To extend the 
accessible $z_{T}$ range, hadrons from $0.5<p_{T}<7.0$ GeV/$c$ are 
used in combination with a single $5<p^{\gamma}_{T}<9$ GeV/$c$ 
photon bin.

Figure~\ref{fig:dphifig1} shows azimuthal pair angle distributions 
for the extracted direct $\gamma$-$h$ correlations in 0--40\% 
central Au$+$Au collisions as well as comparison with the direct 
$\gamma$-$h$ correlations in $p$$+$$p$. Unlike 
on the away-side, on the trigger side ($|\Delta\phi| < \pi/2 $) the 
direct $\gamma$-$h$ correlations in Au$+$Au show a negligible yield, 
indicating that the statistical subtraction method indeed yields 
direct photons and that the yield of fragmentation photons in 
Au$+$Au is negligible within uncertainties.

\begin{figure}
\centering
\includegraphics[width=1.0\linewidth]{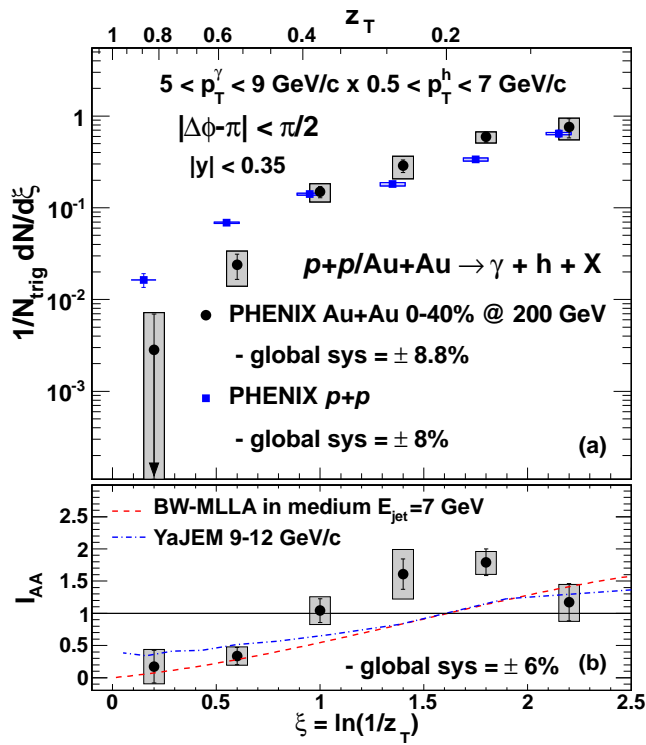}
\caption{(Color online)
The top panel shows per trigger yield as a function of $\xi$ for 
$p$$+$$p$ collisions (squares) and 0--40\% most central Au$+$Au 
collisions (circles). The points are shifted for clarity. For 
reference, the dependence on $z_{T}$ is also indicated. The bottom 
panel shows $I_{\rm AA}$, the ratio of Au$+$Au to $p$$+$$p$ 
fragmentation functions.  Also shown are predictions from 
BW-MLLA~\cite{mlla} (dashed line), calculated at $E_{\rm 
jet}=7$~GeV with $f_{med}=0.8$ selected for 0--10\% central Au+Au 
and from YaJEM-DE~\cite{yajemde,renkprivate} (dot-dashed curve) for 
0--40\% centrality and trigger photons from 9--12~GeV/$c$, both for 
the full away-side ($|\Delta\phi-\pi|<\pi/2$).
}

\label{fig:xidistrib}
\end{figure}

On the away side the associated particle yield is visible, and 
there is significant variation when comparing the correlations in 
Au$+$Au to $p$$+$$p$. To further quantify this variation, the 
yields are integrated over $\Delta\phi$ for $|\pi-\Delta\phi| < 
\pi/2$, as a function of $\xi$, to obtain the effective 
fragmentation function. The top panel of Fig.~\ref{fig:xidistrib} 
shows the integrated away-side yields in Au$+$Au and $p$$+$$p$ as 
circles and squares, respectively. The statistical error bars 
include the point-to-point uncorrelated systematic uncertainty from 
the background subtraction, while the boxes around the points show 
the correlated uncertainties. For reference, the dependence on 
$z_{T}$ is also indicated as the upper scale axis label.

To study medium modification of the jet fragmentation function, 
we take a ratio of the $\xi$ distribution in Au$+$Au to 
$p$$+$$p$. This ratio, known as $I_{\rm AA}$, is shown in the 
bottom panel of Fig.~\ref{fig:xidistrib} and can be written as 
$I_{\rm AA}=Y^{\rm Au+Au}/Y^{p+p}$. Much of the global scale 
uncertainty cancels in this ratio, but there is a remaining 6\% 
uncertainty. In the absence of modification, $I_{\rm AA}$ would 
equal 1. The data instead indicate suppression at low $\xi$ and 
enhancement at higher $\xi$. Including all systematic 
uncertainties the $\chi^2/{\rm dof}$ value for the highest 4 
points compared to the hypothesis that $I_{\rm AA}$ = 1 is 
17.6/4, corresponding to a probability that $I_{\rm AA}$ is 1.0 
for $\xi>0.8$ of less than 0.1\%.

The dashed curve in the bottom panel of Fig.~\ref{fig:xidistrib} 
shows $I_{\rm AA}$ calculated at $E_{\rm jet}=7$~GeV using the 
BW-MLLA model in medium and in vacuum.  The vacuum calculation 
agrees well with the measured $\xi$ distribution in $e^{+}e^{-}$, 
and the in-medium conditions reproduce the measured $\pi^{0}$ 
$R_{AA}$ at high-$p_{T}$ for 0--10\% central Au+Au 
events~\cite{mlla}.  The dot-dashed curve shows $I_{\rm AA}$ 
predicted by YaJEM-DE~\cite{yajemde} for trigger photons from 
9--12~GeV/$c$ for the same centrality range (0--40\%) as the 
present data~\cite{renkprivate}.  Both models, which include all 
away-side jet fragments, show suppression at low $\xi$ due to 
parton energy loss in Au$+$Au collisions, and increasing $I_{\rm 
AA}$ with increasing $\xi$.  In both cases, this is due to the lost 
energy being redistributed into enhanced production of lower 
momentum particles.

\begin{figure}
\centering
\includegraphics[width=1.0\linewidth]{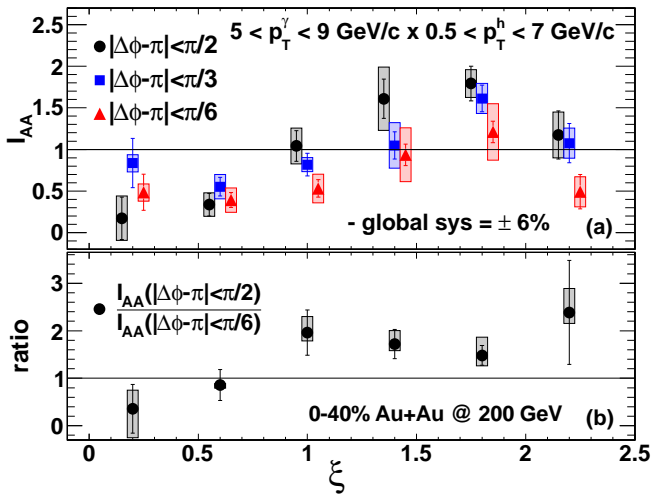}
\caption{(Color online)
The top panel shows the $I_{\rm AA}$ for the full away-side 
($|\Delta\phi-\pi|<\pi/2$) (circles) and for two restricted 
away-side integration ranges, $|\Delta\phi-\pi|<\pi/3$ (squares) 
and $|\Delta\phi-\pi|<\pi/6$ (triangles). The points are shifted 
for clarity. The bottom panel shows the ratio of the $I_{\rm AA}$ 
for $|\Delta\phi-\pi|<\pi/2$ to $|\Delta\phi-\pi|<\pi/6$.}
\label{fig:xiintegrals}
\end{figure}

The suppression of $I_{\rm AA}$ at high $z_{T}$ and enhancement at 
low $z_{T}$ seen in these models agrees with the qualitative trend 
in the data. However, the models do not reproduce the location in 
$\xi$ where transition from suppression to enhancement is observed. 
Understanding the details of this transition can lead to better 
understanding of how lost energy is being redistributed.  One such 
detail is how $I_{\rm AA}$ depends on the angular distribution of 
particles about the away-side jet axis.  The top panel of 
Fig.~\ref{fig:xiintegrals} shows $I_{\rm AA}$ in three integration 
ranges.  Reducing the integration range from $|\Delta\phi-\pi| < 
\pi/2$ reduces the observed enhancement and shifts the effect to 
higher $\xi$.  If the integration range is restricted to 
$|\Delta\phi-\pi| < \pi/6$, the enhancement for $\xi>1.0$ becomes 
negligible, while still showing significant suppression for 
$\xi<0.8$.

To better quantify the angular range of the enhancement, we can 
look at the ratio of $I_{\rm AA}$'s with different integration 
ranges, where some of the statistical and systematic uncertainties 
common to all $I_{\rm AA}$ cancel. The bottom panel of 
Fig.~\ref{fig:xiintegrals} shows the ratio of the full away-side 
integration range to the $|\Delta\phi-\pi| < \pi/6$ case. From this 
ratio it is clear that there is a significant variation in observed 
$I_{\rm AA}$ as a function of the integration range. The average 
ratio for $\xi > 0.8$ is $1.9 \pm 0.3$(stat) $\pm 0.3$(syst), 
indicating that the enhancement in $I_{\rm AA}$ seen at large $\xi$ 
is predominately at large angles ($|\Delta\phi-\pi| > \pi/6$).


In summary, we have presented evidence for medium modification of 
jet fragmentation, measured via comparison of direct photon-hadron 
correlations in $\sqrt{s_{_{NN}}}$ = 200 GeV Au$+$Au and $p$$+$$p$ 
collisions. The ratio of Au$+$Au to $p$$+$$p$ yields indicates that 
particles are depleted at low $\xi$ or high momentum fraction, 
$z_{T}$, due to energy loss of quarks traversing the medium. The 
ratio exhibits an increasing trend toward high $\xi$, exceeding one 
at $\xi \ge 1.0$. Restricting the away-side azimuthal integration range 
reduces the enhancement at high $\xi$ significantly. This suggests 
that the medium enhances production of soft particles in parton 
fragmentation, relative to $p$$+$$p$, preferentially at large 
angles.



We thank the staff of the Collider-Accelerator and Physics
Departments at Brookhaven National Laboratory and the staff of
the other PHENIX participating institutions for their vital
contributions.  
We also thank Thorsten Renk for providing unpublished calculations
and for valuable discussions.
We acknowledge support from the Office of Nuclear Physics in the
Office of Science of the Department of Energy,
the National Science Foundation, 
a sponsored research grant from Renaissance Technologies LLC, 
Abilene Christian University Research Council, 
Research Foundation of SUNY, 
and Dean of the College of Arts and Sciences, Vanderbilt University 
(U.S.A),
Ministry of Education, Culture, Sports, Science, and Technology
and the Japan Society for the Promotion of Science (Japan),
Conselho Nacional de Desenvolvimento Cient\'{\i}fico e
Tecnol{\'o}gico and Funda\c c{\~a}o de Amparo {\`a} Pesquisa do
Estado de S{\~a}o Paulo (Brazil),
Natural Science Foundation of China (P.~R.~China),
Ministry of Education, Youth and Sports (Czech Republic),
Centre National de la Recherche Scientifique, Commissariat
{\`a} l'{\'E}nergie Atomique, and Institut National de Physique
Nucl{\'e}aire et de Physique des Particules (France),
Bundesministerium f\"ur Bildung und Forschung, Deutscher
Akademischer Austausch Dienst, and Alexander von Humboldt Stiftung (Germany),
Hungarian National Science Fund, OTKA (Hungary), 
Department of Atomic Energy and Department of Science and Technology (India),
Israel Science Foundation (Israel), 
National Research Foundation and WCU program of the 
Ministry Education Science and Technology (Korea),
Ministry of Education and Science, Russian Academy of Sciences,
Federal Agency of Atomic Energy (Russia),
VR and Wallenberg Foundation (Sweden), 
the U.S. Civilian Research and Development Foundation for the
Independent States of the Former Soviet Union, 
the US-Hungarian Fulbright Foundation for Educational Exchange,
and the US-Israel Binational Science Foundation.



\begin{thebibliography}{26}
\expandafter\ifx\csname natexlab\endcsname\relax\def\natexlab#1{#1}\fi
\expandafter\ifx\csname bibnamefont\endcsname\relax
  \def\bibnamefont#1{#1}\fi
\expandafter\ifx\csname bibfnamefont\endcsname\relax
  \def\bibfnamefont#1{#1}\fi
\expandafter\ifx\csname citenamefont\endcsname\relax
  \def\citenamefont#1{#1}\fi
\expandafter\ifx\csname url\endcsname\relax
  \def\url#1{\texttt{#1}}\fi
\expandafter\ifx\csname urlprefix\endcsname\relax\def\urlprefix{URL }\fi
\providecommand{\bibinfo}[2]{#2}
\providecommand{\eprint}[2][]{\url{#2}}

\bibitem[{\citenamefont{Adcox et~al.}(2005)}]{whitepaper}
\bibinfo{author}{\bibfnamefont{K.}~\bibnamefont{Adcox}} \bibnamefont{et~al.}
  (\bibinfo{collaboration}{PHENIX Collaboration}), \bibinfo{journal}{Nucl.
  Phys. A} \textbf{\bibinfo{volume}{757}}, \bibinfo{pages}{184}
  (\bibinfo{year}{2005}).

\bibitem[{\citenamefont{Adams et~al.}(2005)}]{starwp}
\bibinfo{author}{\bibfnamefont{J.}~\bibnamefont{Adams}} \bibnamefont{et~al.}
  (\bibinfo{collaboration}{STAR Collaboration}), \bibinfo{journal}{Nucl. Phys.
  A} \textbf{\bibinfo{volume}{757}}, \bibinfo{pages}{102}
  (\bibinfo{year}{2005}).

\bibitem[{\citenamefont{Back et~al.}(2005)}]{phoboswp}
\bibinfo{author}{\bibfnamefont{B.~B.} \bibnamefont{Back}} \bibnamefont{et~al.}
  (\bibinfo{collaboration}{PHOBOS Collaboration}), \bibinfo{journal}{Nucl.
  Phys. A} \textbf{\bibinfo{volume}{757}}, \bibinfo{pages}{28}
  (\bibinfo{year}{2005}).

\bibitem[{\citenamefont{Arsene et~al.}(2005)}]{brahmswp}
\bibinfo{author}{\bibfnamefont{I.}~\bibnamefont{Arsene}} \bibnamefont{et~al.}
  (\bibinfo{collaboration}{BRAHMS Collaboration}), \bibinfo{journal}{Nucl.
  Phys. A} \textbf{\bibinfo{volume}{757}}, \bibinfo{pages}{1}
  (\bibinfo{year}{2005}).

\bibitem[{\citenamefont{Adcox et~al.}(2002)}]{ppg003}
\bibinfo{author}{\bibfnamefont{K.}~\bibnamefont{Adcox}} \bibnamefont{et~al.}
  (\bibinfo{collaboration}{PHENIX Collaboration}), \bibinfo{journal}{Phys. Rev.
  Lett.} \textbf{\bibinfo{volume}{88}}, \bibinfo{pages}{022301}
  (\bibinfo{year}{2002}).

\bibitem[{\citenamefont{Adler et~al.}(2002)}]{star}
\bibinfo{author}{\bibfnamefont{C.}~\bibnamefont{Adler}} \bibnamefont{et~al.}
  (\bibinfo{collaboration}{STAR Collaboration}), \bibinfo{journal}{Phys. Rev.
  Lett.} \textbf{\bibinfo{volume}{89}}, \bibinfo{pages}{202301}
  (\bibinfo{year}{2002}).

\bibitem[{\citenamefont{Weidemann}(2010)}]{urse}
\bibinfo{author}{\bibfnamefont{U.}~\bibnamefont{Weidemann}},
  \emph{\bibinfo{title}{Relativistic Heavy Ion Physics, Landoldt-Boernstein}},
  vol.~\bibinfo{volume}{23} (\bibinfo{publisher}{Springer-Verlag},
  \bibinfo{year}{2010}).

\bibitem[{\citenamefont{Baier et~al.}(2000)\citenamefont{Baier, Schiff, and
  Zakharov}}]{bdmps}
\bibinfo{author}{\bibfnamefont{R.}~\bibnamefont{Baier}},
  \bibinfo{author}{\bibfnamefont{D.}~\bibnamefont{Schiff}}, \bibnamefont{and}
  \bibinfo{author}{\bibfnamefont{B.}~\bibnamefont{Zakharov}},
  \bibinfo{journal}{Annu. Rev. Nucl. Part. Sci.} \textbf{\bibinfo{volume}{50}},
  \bibinfo{pages}{37} (\bibinfo{year}{2000}).

\bibitem[{\citenamefont{Afanasiev et~al.}(2012)}]{raa}
\bibinfo{author}{\bibfnamefont{S.}~\bibnamefont{Afanasiev}}
  \bibnamefont{et~al.} (\bibinfo{collaboration}{PHENIX Collaboration}),
  \bibinfo{journal}{Phys. Rev. Lett.} \textbf{\bibinfo{volume}{109}},
  \bibinfo{pages}{152302} (\bibinfo{year}{2012}).

\bibitem[{\citenamefont{Wang et~al.}(1996)\citenamefont{Wang, Huang, and
  Sarcevic}}]{wang}
\bibinfo{author}{\bibfnamefont{X.-N.} \bibnamefont{Wang}},
  \bibinfo{author}{\bibfnamefont{Z.}~\bibnamefont{Huang}}, \bibnamefont{and}
  \bibinfo{author}{\bibfnamefont{I.}~\bibnamefont{Sarcevic}},
  \bibinfo{journal}{Phys. Rev. Lett.} \textbf{\bibinfo{volume}{77}},
  \bibinfo{pages}{231} (\bibinfo{year}{1996}).

\bibitem[{\citenamefont{Adare et~al.}(2010)}]{ppg095}
\bibinfo{author}{\bibfnamefont{A.}~\bibnamefont{Adare}} \bibnamefont{et~al.}
  (\bibinfo{collaboration}{PHENIX Collaboration}), \bibinfo{journal}{Phys. Rev.
  D} \textbf{\bibinfo{volume}{82}}, \bibinfo{pages}{072001}
  (\bibinfo{year}{2010}).

\bibitem[{\citenamefont{Qin et~al.}(2009)\citenamefont{Qin, Ruppert, Gale,
  Jeon, and Moore}}]{qin}
\bibinfo{author}{\bibfnamefont{G.~Y.}~\bibnamefont{Qin}},
  \bibinfo{author}{\bibfnamefont{J.}~\bibnamefont{Ruppert}},
  \bibinfo{author}{\bibfnamefont{C.}~\bibnamefont{Gale}},
  \bibinfo{author}{\bibfnamefont{S.}~\bibnamefont{Jeon}}, \bibnamefont{and}
  \bibinfo{author}{\bibfnamefont{G.~D.} \bibnamefont{Moore}},
  \bibinfo{journal}{Phys. Rev. C} \textbf{\bibinfo{volume}{80}},
  \bibinfo{pages}{054909} (\bibinfo{year}{2009}).
\bibitem[{\citenamefont{Adler et~al.}(2006{\natexlab{a}})}]{ppg039}
\bibinfo{author}{\bibfnamefont{S.~S.} \bibnamefont{Adler}} \bibnamefont{et~al.}
  (\bibinfo{collaboration}{PHENIX Collaboration}), \bibinfo{journal}{Phys. Rev.
  C} \textbf{\bibinfo{volume}{73}}, \bibinfo{pages}{054903}
  (\bibinfo{year}{2006}{\natexlab{a}}).

\bibitem[{\citenamefont{Adler et~al.}(2006{\natexlab{b}})}]{ppg029}
\bibinfo{author}{\bibfnamefont{S.~S.} \bibnamefont{Adler}} \bibnamefont{et~al.}
  (\bibinfo{collaboration}{PHENIX Collaboration}), \bibinfo{journal}{Phys. Rev.
  D} \textbf{\bibinfo{volume}{74}}, \bibinfo{pages}{072002}
  (\bibinfo{year}{2006}{\natexlab{b}}).

\bibitem[{\citenamefont{Zhang et~al.}(2009)\citenamefont{Zhang, Owens, Wang,
  and Wang}}]{zoww}
\bibinfo{author}{\bibfnamefont{H.}~\bibnamefont{Zhang}},
  \bibinfo{author}{\bibfnamefont{J.~F.}~\bibnamefont{Owens}},
  \bibinfo{author}{\bibfnamefont{E.}~\bibnamefont{Wang}}, \bibnamefont{and}
  \bibinfo{author}{\bibfnamefont{X.~N.} \bibnamefont{Wang}},
  \bibinfo{journal}{Phys. Rev. Lett.} \textbf{\bibinfo{volume}{103}},
  \bibinfo{pages}{032302} (\bibinfo{year}{2009}).

\bibitem[{\citenamefont{Borghini and Wiedemann}()}]{mlla}
\bibinfo{author}{\bibfnamefont{N.}~\bibnamefont{Borghini}} \bibnamefont{and}
  \bibinfo{author}{\bibfnamefont{U.}~\bibnamefont{Wiedemann}},
  \bibinfo{note}{hep-ph/0506218 (2005)}.

\bibitem[{\citenamefont{Renk}(2009)}]{renk}
\bibinfo{author}{\bibfnamefont{T.}~\bibnamefont{Renk}}, \bibinfo{journal}{Phys.
  Rev. C} \textbf{\bibinfo{volume}{80}}, \bibinfo{pages}{014901}
  (\bibinfo{year}{2009}).

\bibitem[{\citenamefont{Adare et~al.}(2009)}]{ppg090}
\bibinfo{author}{\bibfnamefont{A.}~\bibnamefont{Adare}} \bibnamefont{et~al.}
  (\bibinfo{collaboration}{PHENIX Collaboration}), \bibinfo{journal}{Phys. Rev.
  C} \textbf{\bibinfo{volume}{80}}, \bibinfo{pages}{024908}
  (\bibinfo{year}{2009}).

\bibitem[{\citenamefont{Abelev et~al.}(2010)}]{stargjet}
\bibinfo{author}{\bibfnamefont{B.}~\bibnamefont{Abelev}} \bibnamefont{et~al.}
  (\bibinfo{collaboration}{STAR Collaboration}), \bibinfo{journal}{Phys. Rev.
  C} \textbf{\bibinfo{volume}{82}}, \bibinfo{pages}{034909}
  (\bibinfo{year}{2010}).

\bibitem[{\citenamefont{Adcox et~al.}(2003{\natexlab{a}})}]{mainNIM}
\bibinfo{author}{\bibfnamefont{K.}~\bibnamefont{Adcox}} \bibnamefont{et~al.}
  (\bibinfo{collaboration}{PHENIX Collaboration}), \bibinfo{journal}{Nucl.
  Instrum. Methods} \textbf{\bibinfo{volume}{A499}}, \bibinfo{pages}{469}
  (\bibinfo{year}{2003}{\natexlab{a}}).

\bibitem[{\citenamefont{Aphecetche et~al.}(2003)}]{emcNIM}
\bibinfo{author}{\bibfnamefont{L.}~\bibnamefont{Aphecetche}}
  \bibnamefont{et~al.} (\bibinfo{collaboration}{PHENIX Collaboration}),
  \bibinfo{journal}{Nucl. Instrum. Methods} \textbf{\bibinfo{volume}{A499}},
  \bibinfo{pages}{521} (\bibinfo{year}{2003}).

\bibitem[{\citenamefont{Adcox et~al.}(2003{\natexlab{b}})}]{trackingNIM}
\bibinfo{author}{\bibfnamefont{K.}~\bibnamefont{Adcox}} \bibnamefont{et~al.}
  (\bibinfo{collaboration}{PHENIX Collaboration}), \bibinfo{journal}{Nucl.
  Instrum. Methods} \textbf{\bibinfo{volume}{A499}}, \bibinfo{pages}{489}
  (\bibinfo{year}{2003}{\natexlab{b}}).

\bibitem[{\citenamefont{Adare et~al.}(2011)}]{ppg132}
\bibinfo{author}{\bibfnamefont{A.}~\bibnamefont{Adare}} \bibnamefont{et~al.}
  (\bibinfo{collaboration}{PHENIX Collaboration}), \bibinfo{journal}{Phys. Rev.
  Lett.} \textbf{\bibinfo{volume}{107}}, \bibinfo{pages}{252301}
  (\bibinfo{year}{2011}).

\bibitem[{\citenamefont{Sickles et~al.}(2010)\citenamefont{Sickles, McCumber,
  and Adare}}]{ABS}
\bibinfo{author}{\bibfnamefont{A.}~\bibnamefont{Sickles}},
  \bibinfo{author}{\bibfnamefont{M.~P.} \bibnamefont{McCumber}},
  \bibnamefont{and} \bibinfo{author}{\bibfnamefont{A.}~\bibnamefont{Adare}},
  \bibinfo{journal}{Phys. Rev. C} \textbf{\bibinfo{volume}{81}},
  \bibinfo{pages}{014908} (\bibinfo{year}{2010}).

\bibitem[{\citenamefont{Renk}(2011)}]{yajemde}
\bibinfo{author}{\bibfnamefont{T.}~\bibnamefont{Renk}}, \bibinfo{journal}{Phys.
  Rev. C} \textbf{\bibinfo{volume}{84}}, \bibinfo{pages}{067902}
  (\bibinfo{year}{2011}).

\bibitem[{\citenamefont{Renk}()}]{renkprivate}
\bibinfo{author}{\bibfnamefont{T.}~\bibnamefont{Renk}}, \bibinfo{note}{private
  communication}.

\end{thebibliography}

\end{document}